\documentclass{aa}  
\usepackage{graphicx}
\usepackage{txfonts}
\newcommand{\be}{\begin{equation}}
\newcommand{\ee}{\end{equation}}
\newcommand{\lb}[1]{\label{#1}}
\newcommand{\sty}{\scriptstyle}
\newcommand{\ssty}{\scriptscriptstyle}

\newcommand{\apl}{\:^{<}_{\sim}\:}

\newcommand{\dl}{d_{\ssty L}}
\newcommand{\da}{d_{\ssty A}}
\newcommand{\dg}{d_{\ssty G}}
\newcommand{\dz}{d_{\ssty Z}}
\newcommand{\an}{\left[ n \right]}

\newcommand{\oito}{\! \! \! \! \! \! \! \!}
\newcommand{\onze}{\! \! \! \! \! \! \! \! \! \! \!}
\newcommand{\dezessete}{\! \! \! \! \! \! \! \! \! \! \! \! \! \! \! \! \!}
\newcommand{\ud}{^{^{\onze \ssty D}}}
\newcommand{\exud} {^{^{\dezessete \ssty D}}}
\newcommand{\Nid}{\; N_i\ud}
\newcommand{\nid}[1]{\; N_{\ssty {#1}}^{^{\onze \ssty D}}}
\newcommand{\anid}{\; \an_i\exud}
\newcommand{\gammaid}{\; \: \gamma_i^{{\oito \ssty D}}} 
\newcommand{\gestid}{\; \: {\gamma_i^\ast}^{{\onze \ssty D}}\; \; \: \:}
\newcommand{\gamid}[2]{\; \: \gamma_{\ssty #1}^{{\oito \ssty #2}}} 
\newcommand{\gamsid}[2]{\;\:{\gamma_{\ssty #1}^\ast}^{{\onze\ssty #2}}\;\;\:\:}
\newcommand{\NID}[2]{\; N_{\ssty {#1}}^{^{\onze \ssty #2}}}
\begin{document}
   \title{Spatial and observational homogeneities of the
          galaxy distribution in standard cosmologies}
   \titlerunning{Types of homogeneities in standard cosmologies}
%  \subtitle{}
   \author{L. J. Rangel Lemos\thanks{\emph{Present address:} %ICRA,
          International Center for Relativistic Astrophysics, Dept.\
	  of Physics, University of Rome ``La Sapienza'',
          Rome, Italy
          }
          \inst{1}
          \and
          M. B. Ribeiro\inst{2}
          }
   \offprints{M. B. Ribeiro}
   \institute{Department of Astronomy, Valongo Observatory, University
              of Brazil -- UFRJ, Rio de Janeiro, Brazil\\
         \and
             Physics Institute, University of Brazil -- UFRJ,
	     CxP 68532, Rio de Janeiro, RJ 21941-972, Brazil\\
             \email{mbr@if.ufrj.br}
             }
   \date{Received ; accepted }
% \abstract{}{}{}{}{} 
% 5 {} token are mandatory
  \abstract
%  {Empirical verification of the geometrical concept of homogeneity
%  of the standard relativistic cosmology considering its various
%  definitions of distance and emphasizing that astronomical
%  observations occur along the past light cone.}
   {An important aim of standard relativistic cosmology is the empirical
   verification of its geometrical concept of homogeneity by
   considering various definitions of distance and astronomical
   observations occurring along the past light cone.}
   {We analyze the physical consequences of distinguishing between
    \emph{spatial homogeneity} (SH), defined by the Cosmological
    Principle, and \emph{observational homogeneity} (OH). We argue
    that OH is falsifiable by means of astronomical observations,
    whereas SH can be verified only indirectly.}
   {We simulate observational counts of cosmological sources, such as
   galaxies, by means of a generalized number-distance expression
   that can be specialized to produce either the counts of the Einstein-de
   Sitter (EdS) cosmology, which has SH by construction, or other types
   of counts, which do, or do not, have OH by construction. Expressions
   for observational volumes are derived using the various
   cosmological-distance definitions in the EdS cosmological model. The
   observational volumes and simulated counts are then used to derive
   differential densities. We present the behavior of
   these densities for increasing redshift values.}
   {Simulated counts that have OH by construction do not always exhibit
   SH features. The reverse situation is also true. In addition,
   simulated counts with no OH features at low redshift begin to show OH
   characteristics at high redshift. The comoving distance appears to be
   the only distance definition for which both SH and OH are applicable
   simultaneously, even though with limitations.}
   {We demonstrate that observations indicative of a possible absence
   of OH do \emph{not} necessarily falsify the standard Friedmannian
   cosmology, which implies that this cosmology does not always
   produce observable homogeneous densities. We conclude that using
   different cosmological distances in the characterization of the
   galaxy distribution can produce significant ambiguities in reaching
   conclusions about the large-scale galaxy distribution in the Universe.}

   \maketitle
%________________________________________________________________

\section{Introduction}

The determination of whether or not the large-scale distribution
of matter in the Universe reaches a homogeneous distribution on
some redshift scale has been a disputed topic in observational
cosmology for many decades. With the availability of increasingly
larger galaxy redshift survey databases stemming from increasingly
more complete and deeper galaxy samples, there have been renewed
efforts to solve this problem by means of statistical techniques
of growing sophistication. However, this issue has still not been
settled because contradictory results have been reported by various
authors who support opposite claims (Ribeiro \& Miguelote 1998;
Sylos Labini et al.\ 1998; Joyce et al.\ 1999, 2000, 2005; Gabrielli
et al.\ 2005; Hogg et al.\ 2005; Jones et al.\ 2005; Yadav et al.\
2005).  

The conventional wisdom on this topic claims that the solution to
the problem lies in our ability to acquire more and better data, that
is, as more complete galaxy samples at both lower and higher redshift
ranges become available the statistical techniques currently applied
to the analysis of these data should suffice to settle the controversy. 
In other words, this view assumes implicitly or explicitly that more
complete galaxy redshift survey samples will eventually clarify this
point once and for all. The difficulty with the conventional wisdom is
that we have witnessed an enormous improvement in the technological
methods for astronomical data acquisition and, as a direct result of
those technological advances, have indeed obtained higher quality
and more complete galaxy redshift survey data sets. Despite these
technical advances, the controversy, however, continues to resurface
(see Ribeiro 1994 for a brief historical account of this debate in the
past century). Ribeiro (1992) proposed that the source of the 
controversy was not the observations themselves, but the conceptual
tools used to analyse the observations. The initial ideas were
gradually refined and an alternative perspective developed (Ribeiro
2001b, 2005; Albani et al.\ 2007). This perspective suggests that
improvements in observational techniques and the acquisition of larger
and larger galaxy data samples is not the path that will shed light
and clarify this debate.

This alternative conceptual framework grew out of various 
works (Ribeiro 1992ab, 1993, 1994, 1995, 2001a; Abdalla et al.\
2001) and was comprehensively analyzed in Ribeiro (2001b; hereafter
R01b). Albani et al.\ (2007; hereafter A07) further developed this
alternative perspective in a somewhat condensed version. First
principles were used to point out that General Relativity allows us
to define two different concepts of homogeneity perfectly applicable
to cosmological models: \emph{spatial homogeneity} (SH) and
\emph{observational homogeneity} (OH). Furthermore, it was argued
that the Cosmological Principle is based on the concept of SH,
whereas the astronomical search for the possible homogeneity
of the Universe occurs mostly in the context of OH: this is
because astronomical observations are completed where OH is defined,
that is, along the backward null cone. The Cosmological Principle
implies that SH is not directly observable on space-like surfaces of
constant time defined in the Friedmann-Lema\^{\i}tre-Robertson-Walker
(FLRW) cosmologies. On the other hand, the geometrical locus of OH is
not along those space-like surfaces of constant time, but along the
past null cone, and so OH will only occur if a density measured
directly from observations, usually galaxy number counts, remains
constant along these null surfaces. These two concepts of homogeneity
do overlap, but they are \emph{not} the same.\footnote{In earlier works
(Ribeiro 1992ab, 1993, 1994, 1995), the term ``apparent homogeneity'' was
used instead of the currently adopted term ``observational
homogeneity.''} Reports of the searches for the homogenization of the
matter distribution in the Universe have not, for the most part,
acknowledged this important difference since the use of the generic, but
ambiguous, term `homogeneity' has become commonplace (see Section
\ref{rel} below). As discussed in R01b and A07, such a distinction arises
only if one takes an entirely relativistic perspective of this problem. 

Nevertheless, acknowledging the existence of these two types of
homogeneities is not enough to clarify the controversial points
outlined above. One has to go a step further because the only
way to discriminate SH from OH is by building observational
densities using different distance measures with the same number
count data. Bearing in mind the conceptual framework summarized
above, R01b and Ribeiro (2005; hereafter R05) showed that
although the Einstein-de Sitter (EdS) cosmology has SH by
construction, it may, or may not exhibit OH, since the possible
presence of OH depends on the distance measure chosen
in the statistical analysis of its EdS theoretically derived
number count expression as a function of the redshift. Such a
result was confirmed by A07 using observations, for instance, by
means of number counts extracted from the Canadian
Network Observational Cosmology 2 (CNOC2) galaxy redshift survey
and applied to two types of standard cosmologies, EdS and FLRW
with $\Omega_{m_0} =0.3$, $\Omega_{\Lambda_0}=0.7$.

When these two types of homogeneities are discussed, a question
which also arises is how deep the observations must be to be able
to distinguish SH from OH. In other words, we must determine 
whether or not the redshift ranges of current galaxy surveys
are sufficiently deep to be able to detect this difference. It is
important to point out that previous work
(Ribeiro 1995) indicated that due to the high nonlinearity of
General Relativity the distinction between SH and OH effects
may occur, in theory, at redshifts as low as $z \apl 0.1$,
depending on the chosen relativistic cosmological model and the
observational quantity under study.

The aim of this paper is to analyze further the issues discussed
above. Our goal here is to extend the studies presented in R01b,
R05, and A07. These papers started with SH by construction and
sought to show whether or not OH was also featured in the models.
R01b and R05 initiated their analysis from the theoretical EdS
number counts, whereas A07 used observed number counts
extracted from the luminosity function. We aim here to investigate
the opposite situation, i.e., start with OH by construction and then
investigate whether or not the models show SH. Instead of using
actual observations, our intention here is to
\textit{simulate observations} by means of a generalized
number-distance relation which may, or may not, produce OH from the
start and then search for possible SH in the resulting model.
{In addition, we shall analyze our results in two redshift
ranges, namely $0.001<z<0.1$ and $z>0.1$, since most direct
measurements of galaxy correlations have been limited to
$z \sim 0.1$. This aims to try to answer the question posed above
about the redshift depth which these effects begin to manifest
themselves, as well as probing which quantities could possibly
offer observational results from which we can attempt to obtain
observational evidence allowing us to discriminate between SH and
OH.}\footnote{It should be mentioned here that if one uses
the galaxy luminosity function data derived from various 
surveys with a relativistic cosmology number count theory (Ribeiro
\& Stoeger 2003), one is able to indirectly obtain measurements of
galaxy number counts at $z \approx 1$, or at far higher
redshifts, where the distinction between SH and OH is easily
detected. See A07 and Iribarrem, Ribeiro \& Stoeger (2008, in
preparation).}

Our results show that a model with no OH by construction does not
always remain that way at higher values of redshift. In fact,
with a specific distance definition we may have a model with OH
at low redshifts, but no OH at higher redshifts. We also found
that if we start a model with OH, it may or may not become SH.
These results imply that the use of different distance measures to
calculate cosmological densities produces significant ambiguities
in reaching conclusions about the behavior of the large-scale
distribution of galaxies in the Universe due to the impossibility
of uniquely characterizing densities from galaxy distribution
data. Conclusions on this matter reached by means of the use of just
one cosmological distance, usually comoving, should therefore be
seen as applicable to this distance measure only and are most likely
not valid generally. Therefore, the proposal of R05 that
\textit{observers should utilize all possible distance measures in
their data analysis} is reinforced here. It is the view of these authors
that \textit{this is the path towards clarifying the controversy
discussed above}.

The plan of the paper is as follows. In Sect., \ref{sec2} we derive
the basic equations and definitions of observational distances,
areas, volumes, number counts, and densities in cosmology.
Section \ref{eds} reviews the results of R01b, R05, and A07,
where those definitions are applied to FLRW cosmological models,
this shows that although the EdS cosmology is spatially homogeneous
by construction, it has observational homogeneity only when the
comoving distance is adopted in calculating observational
densities. In Sect., \ref{sec4} we simulate models that do, or do
not, have observationally-homogeneous features, concluding
that even a model without OH at low redshifts may become
observationally homogeneous at higher values of $z$. Section
\ref{DSH} studies the asymptotic behavior of the fractal
dimension $D$ of the EdS model at $z\rightarrow 0$ and at the
Big Bang singularity hypersurface where $z \rightarrow \infty$,
showing that the former case leads to $D=3$ for all distance
measures. In contrast, the latter case implies that $D=0$ for all
distances, except the comoving distance where the value
$D=3$ remains unchanged for all $z$. { Section \ref{mags} discusses the
relationship between number counts $N$ and magnitudes by means
of the distance modulus $\mu$. We show that the same ambiguous
results obtained in Sect.\ \ref{sec4} for the $N \times z$
functions, constructed with the various distance definitions,
are also present in the $N \times \mu$ functions.} Section
\ref{rel} provides a conceptual discussion about the caveats of
the use of the generic term `homogeneity' in cosmological models,
arguing that observational homogeneity is a relative concept
entailed by the relativity of time intervals. Finally, Sect.\
\ref{conclu} summarizes the results obtained in this paper.

%__________________________________________________________________

\section{Distances, volumes, densities and number counts}\lb{sec2}

As has been extensively argued elsewhere, measuring distances in
cosmology depends on circumstances, that is, on the method of
measurement (McVittie 1974, Sandage 1988, R01b, R05 and references
therein). This does not imply that distances cannot be
compared with each another. These are true, physical distances to
an object and they can indeed be compared simply because they are
distances to cosmological sources, mostly galaxies, for which
intrinsic physical characteristics can be determined
(intrinsic measurement, intrinsic luminosity, etc)
\emph{independently} of a cosmological model. In addition, the
reciprocity theorem (Ellis 2007) relates the various
distances to each another and allows conversions between them.

That is the theory. In practice, however, due to technological
limitations and our incomplete knowledge of the physical processes
occurring in the evolution of galaxies, we are presently unable
to find those intrinsic measurements, that is, standard candles and
standard rods, for \emph{every} galaxy in a redshift survey.
Therefore, we are left to measure their redshifts only and, by
using a cosmological model, to relate those redshifts to some
distance. Textbooks and reviews discussing cosmological distance
definitions offer a plethora of names for the distance measures,
a fact which only adds confusion to, not infrequently, poorly
understood concepts about what is a distance to a cosmological
object, the definition that should be chosen, and the context for
choosing a certain definition and not another one. As soon as one
delves into this problem and reads how it is dealt with in the
literature, it becomes clear that familiar Newtonian concepts
slipped into a subject that can only be understood properly by
means of relativistic ideas. Thus, to avoid those Newtonian
concepts of absolute and unique definitions, which are not applicable
to the relativistic discussion proposed here, we must follow along
the wise footsteps of others and accept that \textit{there is no
such a thing as an unique cosmological distance: all are correct,
and all can be compared with each another.} To argue otherwise is to
allow Newtonian ideas to slip into a subject that is entirely
relativistic.

The difficulties described above have, of course, been previously
perceived by others, such as observational cosmologists, who resorted
to the convenient convention of using, for
the most part, only one distance measure, the \emph{comoving
distance}. There are, however, three caveats to this practice.
Firstly, not all practitioners follow this convention and this
means that there are still those who are misled into treating
different distance definitions as if they were the same, when
they are not, and, worst of all, results derived from those
different distance measures are then improperly compared with each
other. This can obviously add even more confusion to an
already-confused subject. As we show below, the second caveat is
that the comoving distance implies that the task of distinguishing
spatial from observational homogeneity becomes very difficult.
Thirdly, by adopting just one distance as convention some may still
unconsciously fall into the familiar, but in this context very
misleading, \textit{Newtonian trap of believing that in cosmology
the distance to an object could be uniquely defined, when General
Relativity does not allow such a conclusion.}

\subsection{Cosmological distances}

Our proposed solution to these difficulties is therefore to use in
our analysis below all observational distances, the \emph{luminosity
distance} $\dl$, the \emph{area distance} $\da$, the \emph{galaxy
area distance} $\dg$, and the \emph{redshift distance}
$\dz$.\footnote{$\da$ is also known as `angular diameter distance',
`corrected luminosity distance' and `observer area distance'.
$\dg$ is also known as `effective distance', `angular size
distance', `transverse comoving distance' and `proper motion
distance' (see details and references in R01b and R05). It is
also possible to define another distance, the \textit{parallax
distance} $d_{\ssty P}$ due to galaxy parallaxes (Ellis 1971).
This distance is not often mentioned since galaxy parallaxes
cannot yet be measured.} These are quantities that can in principle,
be directly measured (R05; Ellis 2007). However, since we need to
assume a cosmological model to start with, other distances like the
comoving distance, proper distance, etc, can all be written in
terms of these four above (see below). The advantage of starting
with these observational four, which to simplify the notation
from now on shall be referred as $d_i \; (i={\sty A}, {\sty G},
{\sty L}, {\sty Z} \,)$, is to know beforehand that they are
defined along the past light cone, since they are observational
distances. Relating them to, for example, comoving distance, implies
that the solution of the null geodesic equation has therefore to be
included in the expression of the comoving distance. This is
standard practice, rarely mentioned, but, for the purposes of
this work, it must be stated explicitly to avoid confusion.

These observational distances are related to each other by an
important result called the \textit{ reciprocity theorem},
or \textit{Etherington reciprocity law}, proven long ago by
Etherington (1933), which reads as follows,\footnote{
See Ellis (2007) for an appraisal of how fundamental this
theorem is in every aspect of modern cosmology.}
\be {(1+z)}^2 \da=(1+z) \: \dg = \dl.
    \lb{eth}
\ee
This theorem is valid for \textit{all} cosmologies (Ellis 1971,
2007; Pleba\'{n}ski and Krasi\'{n}ski 2006). In addition, R01b
and R05 also defined a distance by redshift as follows,
\be \dz=\frac{cz}{H_0},
    \lb{dz}
\ee
where $c$ is the light speed and $H_0$ is the Hubble constant.
This is not a distance in the sense of the other distance measures:
since it is often used in observational cosmology, it is, however,
useful to adopt it here and assume Eq.\ (\ref{dz}) to be the definition
of redshift distance $\dz$ for all $z$.

\subsection{Observational areas and volumes}

We now define observational areas and volumes. Following R05,
the area of the observed spherical shell of radius $d_i$ may be
written as,
\be S_i=4 \pi {(d_i)}^2, \lb{area} \ee
and the observational volume at this same radius can be
straightforwardly written as,
\be V_i=\frac{4}{3} \pi \, {(d_i)}^3.
    \lb{vol}
\ee
{These equations imply that the observed volume element
is given by,
\be dV_i=S_i \; d(d_i).
    \lb{thick}
\ee
Here $d(d_i)$ is the elementary shell thickness and, since
cosmological distances are function of the redshift, the shell
thickness can be written approximately as $\Delta d_i = \left[
d(d_i)/dz \right] \Delta z$, for a certain observed redshift
interval $\Delta z$.}

These expressions are completely general and Eq.\ (\ref{vol})
agrees with the usual volume definitions. We note that these areas
and volumes are observational, that is, they are defined along
the past light cone.

%______________________________________________________________

\subsection{Generalized number-distance relation}

We aim to simulate number counts as if they were actual observations. To
do so in a general manner, it is
convenient to adopt the following expression for the \emph{number
of observed cosmological sources} $N$ at a certain observational
distance $d_i$,
\be \Nid = { \left( B \, d_i \right) }^D.
    \lb{nd}
\ee
Here $B$ is an as yet unspecified constant and $D$ is the fractal
dimension (Pietronero 1987; Ribeiro and Miguelote 1998; Sylos
Labini et al.\ 1998). This expression provides a general
way of simulating galaxy counts, since, for $D=3$ we have an
\textit{observationally}-homogeneous distribution (see below),
whereas for $D<3$ we have an \textit{observationally-in}homogeneous
galaxy counting. The subscript index $i$ and superscript index $D$
therefore label respectively the choice of distance and how far the
simulated counting differs from OH. We note that it is  \textit{not}
our intention to either prove or disprove the possible fractality of
the galaxy distribution. The adoption of the equation above for $N$
is just a matter of convenience due to its generality.

de Vaucouleurs (1970) and Wertz (1970, 1971) long ago proposed similar
relations to Eq. (\ref{nd}), which were written instead in terms of
density. They appear to be the first authors to use it in the context
of galaxy distributions. Pietronero (1987) independently advanced an
expression virtually identical to the above. As discussed in Ribeiro
and Miguelote (1998; see also Ribeiro 1994), the models of Wertz
(1970, 1971) and Pietronero (1987) shared more similarities than
differences, both conceptually and analytically. Wertz did not use
the word `fractal', although self-similar ideas can be found in his
discussion. Pietronero (1987) named his expression the ``generalized
mass-length relation''; here we instead choose to refer to Eq.\
(\ref{nd}) as the \emph{generalized Pietronero-Wertz number-distance
relation}, or simply \emph{generalized number-distance relation}, since
we believe the emphasis on number-distance, rather than mass-length,
is more appropriate to observational cosmology. 

%______________________________________________________________

\subsection{Observational densities}

As discussed in R05 and A07, the \textit{differential density}
$\, \! \! \gammaid$ at a certain distance $d_i$, and with a specific
choice of the dimension $D$, is defined by the following expression,
\be \gammaid = \frac{1}{S_i} \frac{d \left( \Nid \right)}{d \left(
               d_i \right) }.
    \lb{diff}
\ee
This equation provides a measure of the rate of growth in the number
density as one moves down the past light cone along the observable
distance $d_i$. Obviously, the behavior of the differential density
depends heavily on the distance employed in Eq.\ 
(\ref{diff}). The \textit{integral differential density} $\gestid$
is defined to be the integration of $\gammaid$ over the observational
volume $V_i$, corresponding to,
\be \gestid= \frac{1}{V_i} \int_{V_i} \gammaid \: dV_i.
    \lb{gest}
\ee
If we now define $\! \anid$ to be the \textit{radial number
density} for a given distance measure $d_i$, the following
result clearly holds, once we consider Eqs.\ (\ref{area}),
(\ref{vol}), (\ref{diff}), and (\ref{gest}),
\be  \anid = \frac{\Nid}{V_i} = \gestid.
    \lb{densi}
\ee

{At this point some important remarks are necessary.
The radial number density $\! \anid$ considers the number
counts of objects as a function of distance from a single point.
In the context of FLRW cosmology, this single point can be any
point in a 4-dimensional Riemannian manifold, since this spacetime
assumes a maximal spatial isotropy (see \S \ref{rel} below).
Therefore, this is not an average quantity, or ensemble average,
obtained as an average made over many realizations of a stationary
stochastic process, where the ensemble average can be replaced by a
volume average. Confusion arises when Eq.\ (\ref{nd}) is viewed 
in the framework of fractal geometry, where the number count of
this equation is interpreted as an average quantity. Here Eq.\
(\ref{nd}) is simply the radial matter distribution which
\textit{could} be given by a perfect fluid approximation of
cosmological solutions of Einstein's field equations. As
discussed in R05, the relationship between these two quantities,
the radial number density and the average number density, remains an
open question although it appears to be 
reasonable to assume that they are related. The present paper
is not concerned with proving or disproving the fractal hypothesis
for the galaxy distribution, but aims to investigate the limitations
of the standard concept of ``homogeneity'', in particular 
its application to interpreting real astronomical observations.}

It is useful to write the differential densities 
in terms of the redshift $z$. The results are as follows,
\be \gammaid (z) = \left[ \frac{d}{dz} \left( \Nid \right) \right]
               {\left[ S_i \frac{d}{dz} \left( d_i \right)
	       \right]}^{-1},
    \lb{gz}
\ee
\be \gestid (z) = \frac{1}{V_i} \int\limits_0^z \gammaid \left(
             \frac{dV_i}{dz'} \right) dz' = \frac{\Nid(z)}{V_i(z)}.
    \lb{gestz}
\ee
In both of the above equations one can clearly identify two distinct parts.
The first term is the geometrical term, determined by the spacetime
geometry of the chosen metric; this is the case of the functions
$d_i(z)$, $S_i(z)$ and $V_i(z)$ and their derivatives. The second
term is given by the number count $\! \Nid(z)$ and the differential
number count $d \left[ \Nid(z) \right] \big/ dz$ and are determined
either by theory or observationally. Various tests of cosmological
models rely, in one way or another, on the comparison of number
counting, determined observationally, with its theoretical prediction.
We note that, in Eq.\ (\ref{gz}), this division between the geometrical
and theoretical/observational parts are clearly visible, since they
are represented by each term inside the brackets on the right hand
side.

As mentioned above, the geometrical part was determined entirely by
assuming special cases of the FLRW metric, EdS cosmology in R01b, R05,
and A07 and the FLRW open model in A07. The number count, however,
was either theoretically determined from the cosmological model, that
is, by taking the expression for $N(z)$ as given by the matter
distribution in these cosmologies, which meant assuming a
spatially-homogeneous matter distribution
from the start (R01b, R05), or by using differential counts
$dN/dz$ obtained from the luminosity function of galaxy surveys,
as demonstrated by A07.\footnote{It must be noted that when comparing
densities constructed from observationally-derived number
counts with the theoretical predictions of a FLRW model with
$\Omega_{m_0} =0.3$, $\Omega_{\Lambda_0}=0.7$, that is, having
SH by construction, A07 found deviations from pure SH. However,
it was not clear if these deviations were due to possible
incompleteness of the sample, the use of an inappropriate
evolution function when deriving the luminosity function
parameters, or if they were true deviations from SH. See details
in the caption Figs.\ 7 and 8 of A07.} We 
determine the geometrical part by choosing a spacetime metric,
since this is unavoidable; instead of restricting ourselves however
to obtaining the number count solely from the chosen cosmology,
we use Eq.\ (\ref{nd}) to simulate number counts, \textit{as if} they
were real observations. This methodology attempts to make possible
the investigation of the model behavior, when OH is assumed from the
outset; it configures an approach that is opposite to that investigated
in R01b, R05, and A07. The advantage of this methodology is to depart
from pure spatially-homogeneous cosmologies. 

%______________________________________________________________

\section{Observational \underline{in}homogeneity of the Einstein-de
         Sitter spacetime}\lb{eds}

We shall assume this spacetime metric for analytical simplicity
only, but this choice does not affect our main results (see below).
The methodology that we present can be extended easily to other
FLRW spacetime metrics. The EdS expressions shown next were derived
in R05 (see also R01b).

In the EdS model, the four observational distances discussed above are
given by the following set of equations,\footnote{In EdS cosmology,
the comoving distance is equal to the galaxy area distance $\dg$ 
multiplied by a constant factor (see R05).} 
\be \da(z)=\frac{2c}{H_0}  \left[ \frac{1+z- \sqrt{1+z}}{{ \left(
           1+z \right) }^2} \right],
    \lb{da}
\ee
\be \dg(z) =\frac{2c}{H_0}  \left( \frac{1+z- \sqrt{1+z}}{1+z}
            \right),
    \lb{dg}
\ee
\be \dl(z) =\frac{2c}{H_0}  \left( 1+z- \sqrt{1+z} \right),
    \lb{dl}
\ee
\be \dz(z)= \frac{2c}{H_0}  \left( \frac{z}{2}  \right).
    \lb{dz1}
\ee
{Equations (\ref{da}), (\ref{dg}), and (\ref{dl}) have
Taylor-series expansions in terms of redshift that are given by,
\be \da(z)=\frac{c}{H_0} \left(z-\frac{7}{4}z^2+\frac{19}{8}
    z^3+ \ldots \right),
    \lb{daz}
\ee
\be \dg(z)=\frac{c}{H_0} \left(z-\frac{3}{4}z^2+\frac{5}{8}
    z^3+ \ldots \right),
    \lb{dgz}
\ee
\be \dl(z)=\frac{c}{H_0} \left(z+\frac{1}{4}z^2-\frac{1}{8}
    z^3+ \ldots \right),
    \lb{dlz}
\ee
which shows that all four distances above reduce to the same expression
to first order.}

The number count in this cosmology is well known, producing the
following expression,
\be N^{\mathrm{\ssty EdS}}= \alpha\; 
    { \left( \frac{1+z-\sqrt{1+z}}{1+z} \right) }^3,
    \lb{neds}
\ee
where the dimensionless constant $\alpha$ is defined to be, 
\be \alpha=\frac{4c^3}{H_0 M_g G}.
    \lb{alpha}
\ee
Here $M_g$ is the average galactic rest mass $( \sim 10^{11}
M_\odot )$ and $G$ is the gravitational constant. The differential
densities in this cosmology are given by the following equations, 
\be \gamma_{\ssty A}^{{\mathrm{\ssty EdS}}} = \mu_0 
    \left[ { \frac{{(1+z)}^3}{(3-2 \sqrt{1+z} ) } } \right] ,
    \lb{ga}
\ee
\be \gamma_{\ssty G}^{{\mathrm{\ssty EdS}}} = \mu_0,
    \lb{gg}
\ee
\be \gamma_{\ssty L}^{{\mathrm{\ssty EdS}}} = \mu_0 
    \left[ \frac{1}{ (2 \sqrt{1+z} - 1) { (1+z)}^3} \right],
    \lb{gl}
\ee
\be \gamma_{\ssty Z}^{{\mathrm{\ssty EdS}}} = \mu_0 \left[ \frac{4 \,
    {(1+z- \sqrt{1+z}) }^2}{z^{\, 2} \, { \left( 1+z \right)
    }^{7/2}} \right],
    \lb{gz1}
\ee
and the integral differential densities yield,
\be \gamma_{\ssty A}^{\ast^{\mathrm{EdS}}} = \mu_0
    { \left(1+z \right) }^{3},
    \lb{gamasa}
\ee
\be \gamma_{\ssty G}^{\ast^{\mathrm{EdS}}} = \mu_0,
    \lb{gamasg}
\ee
\be \gamma_{\ssty L}^{\ast^{\mathrm{EdS}}} = \mu_0 
     { \left( 1+z \right) }^{-3},
    \lb{gamasl}
\ee
\be \gamma_{\ssty Z}^{\ast^{\mathrm{EdS}}} = \mu_0 { \left[
    \frac{ 2(1+z-\sqrt{1+z}) }{z { \left( 1+z \right)
    }} \right] }^3,
    \lb{gamasz}
\ee
where the constant $\mu_0$ is given as below,
\be \mu_0=  \frac{3 {H_0}^2}{8\pi M_g G}.
    \lb{mu0}
\ee 

{Taylor series expansions for the differential densities
above are given as follows,
\be \gamma_{\ssty A}^{{\mathrm{\ssty EdS}}} = \mu_0
    \left(1+4z+\frac{27}{4}z^2+\frac{55}{8}z^3+\ldots \right)
    \lb{gasz}
\ee
\be \gamma_{\ssty L}^{{\mathrm{\ssty EdS}}} = \mu_0 
    \left(1-4z+\frac{41}{4}z^2-\frac{171}{8}z^3+\ldots \right)
    \lb{glsz}
\ee
\be \gamma_{\ssty Z}^{{\mathrm{\ssty EdS}}} = \mu_0  
    \left(1-3z+\frac{95}{16}z^2-\frac{39}{4}z^3+\ldots \right)
    \lb{gz1sz}
\ee
\be \gamma_{\ssty A}^{\ast^{\mathrm{EdS}}} = \mu_0 
    \left(1+3z+3z^2+z^3 \right)
    \lb{gamasasz}
\ee
\be \gamma_{\ssty L}^{\ast^{\mathrm{EdS}}} = \mu_0 
    \left(1-3z+6z^2-10z^3+\ldots \right)
    \lb{gamaslsz}
\ee
\be \gamma_{\ssty Z}^{\ast^{\mathrm{EdS}}} = \mu_0  
    \left(1-\frac{9}{4}z+\frac{57}{16}z^2-\frac{39}{8}z^3+\ldots \right)
    \lb{gamaszsz}
\ee
We note that all densities above have a non-vanishing zeroth order
term, whereas the series expansions for the cosmological distances,
given by Eqs.\ (\ref{daz}), (\ref{dgz}), and (\ref{dlz}), do not contain
this term. This implies that, according to Eqs.\
(\ref{gasz})--(\ref{gamaszsz}), deviations from a constant density
value, that is, from OH (see below), occur in the first-order
terms of the series, whereas Eqs.\ (\ref{daz})--(\ref{dlz})
imply that deviations from the Hubble law occur in the second-order
terms. In other words, Hubble-law deviations occur in higher
redshift ranges than deviations from OH. This was first noticed
in Ribeiro (1995), discussed again in R01b and explored further in
Abdalla et al.\ (2001) by means of a simple perturbed model.} 

As discussed at length in R01b, OH corresponds to a constant value
of observational average density, that is, when this average
density is calculated along the chosen spacetime's past null cone.
This requirement was operationally defined in R05 and A07 to
mean the following condition,
\be \gamma_i^\ast = \mathrm{constant}, \; \; \; \; \; \; \; \;
    \mbox{\rm (OH definition)}.
    \lb{oh}
\ee
It is clear from a simple inspection of Eqs.\ (\ref{gamasa}),
(\ref{gamasg}), (\ref{gamasl}), and (\ref{gamasz}) that although EdS
cosmology is SH by construction for \textit{all} distance measures,
according to the definition provided by Eq.\ (\ref{oh}) the conditions
of OH are met only when one builds an average density using the galaxy
area distance $\dg$ (see Eq.\ \ref{gamasg}). The three other distance
definitions, $\da$, $\dl$, and $\dz$ are apparently unsuitable 
as tools for searching the possible OH in the galaxy redshift data,
if one adopts the EdS cosmology. This was the main conclusion reached
by R05, which was also found to be valid for open FLRW cosmology
in A07. This conclusion reproduced here to compare with
the analysis below is graphically summarized in Figs.\ \ref{fig1}
and \ref{fig12}. 
\begin{figure}
\centering
\vspace{-5.cm}
\hspace{-2.cm} \includegraphics[width=12.5cm]{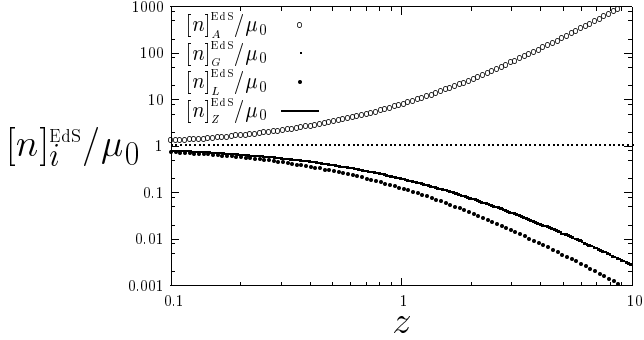}
\vspace{-8.5cm}
\caption{Plot similar to the one appearing in R05 showing the
         normalized radial number densities ${\left[ n \right]}_i$,
	 or integral differential densities $\gamma_i^\ast$ (see
	 eq.\ \ref{densi}), along the past light cone versus the
	 redshift $z$ in the EdS cosmology. Note that although all
	 cases are spatially-homogeneous by construction, only
	 ${\left[ n \right]}_{ \ssty G}^{\mathrm{\ssty EdS}}$ is
	 observationally homogeneous as well. The other three
	 radial number densities are spatially homogeneous, but
	 observationally {\it in}homogeneous. The asymptotic
	 limits for these densities are also different (see R01b),
	 yielding, $\lim\limits_{z \rightarrow \infty}
	 {\left[ n \right]}_{\ssty L}^{\mathrm{\ssty EdS}} = 0$,
         $\lim\limits_{z \rightarrow \infty}
	 {\left[ n \right]}_{\ssty A}^{\mathrm{\ssty EdS}} =\infty$,
         $\lim\limits_{z \rightarrow \infty}
	 {\left[ n \right]}_{\ssty G}^{\mathrm{\ssty EdS}} = \mu_0$,
         $\lim\limits_{z \rightarrow \infty}
	 {\left[ n \right]}_{z}^{\mathrm{\ssty EdS}} = 0$.
         }
    \lb{fig1}
\end{figure}
\begin{figure}
\centering
\vspace{-5.cm}
\hspace{-2.cm} \includegraphics[width=12.5cm]{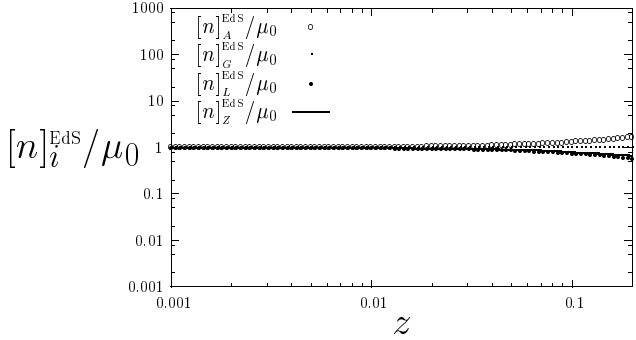}
\vspace{-8.5cm}
\caption{This figure shows the normalized radial number
         densities ${\left[ n \right]}_i$ of the previous graph,
	 but at a redshift range of up to $z=0.2$. As noted in Ribeiro
	 (1995), the distinction between the densities built with
	 different distance measures can be seen before the
	 redshift reaches $z=0.1$.}
    \lb{fig12}
\end{figure}

\section{Models with or without observational homogeneity}\lb{sec4} 

We now adopt the generalized number-distance relation provided by Eq.\
(\ref{nd}) to obtain more general models, whose main properties have or
do not have OH. As discussed above, we shall use Eq.\ (\ref{nd}) to
simulate possible matter distributions and test whether or not an assumed
OH number count distribution produces SH, or if a distribution with no
OH produces, or not, models with no SH.

Considering the definitions given by Eqs.\ (\ref{area}) and
(\ref{vol}), it is straightforward to conclude that the
generalized number-distance relation provided by Eq.\ (\ref{nd})
corresponds to the following expressions for the differential density
given by Eq.\ (\ref{diff}) and the integral differential density
given by Eq.\ (\ref{gest}),
\be \gammaid \, (d_i) = \frac{DB^D}{4\pi} d_i^{D-3}, 
    \lb{diffPW}
\ee
\be \gestid (d_i)= \frac{3B^D}{4\pi} d_i^{D-3}.
    \lb{gestPW}
\ee
From these results, it becomes obvious that these
two densities are related by the following expression,
\be \gestid = \frac{3}{D}\, \gammaid \, .
    \lb{doisgs}
\ee

To proceed with our analysis, it is unavoidable at this stage to
choose a cosmological model (see Sect.\ \ref{sec2} above). Our
choice then is to continue using an EdS cosmology, due to its
simplicity, for everything besides number counts. However, our
results can be extended to other FLRW spacetimes, or even
to non FLRW cosmologies.

We substitute the four EdS distance definitions, given by
Eqs.\ (\ref{da}), (\ref{dg}), (\ref{dl}), and (\ref{dz1}) into
Eq.\ (\ref{nd}). The results may be written as follows,
\be {\; \nid{A}}(z)={\left(\frac{2cB}{H_0}\right)}^D \;
    {\left[\frac{1+z-\sqrt{1+z}}{{(1+z)}^2} \, \right]}^D,
    \lb{Nad}
\ee
\be {\; \nid{G}}(z)={\left(\frac{2cB}{H_0}\right)}^D \;
    {\left(\frac{1+z-\sqrt{1+z}}{1+z} \, \right)}^D,
    \lb{Ngd}
\ee
\be {\; \nid{L}}(z)={\left(\frac{2cB}{H_0}\right)}^D \;
    {\left({1+z-\sqrt{1+z}} \, \right)}^D,
    \lb{Nld}
\ee
\be {\; \nid{Z}}(z)={\left(\frac{2cB}{H_0}\right)}^D \;
    {\left(\frac{z}{2}\right)}^D.
    \lb{Nzd}
\ee
If we now substitute these EdS distance measures into Eq.\
(\ref{diffPW}), we may write the expressions for the differential
densities as shown below,
\be \gamid{A}{D}(z) = \left(\frac{D{H_0}^3}{32\pi c^3}\right)
    {\left(\frac{2cB}{H_0}\right)}^D \;
    {\left[\frac{1+z-\sqrt{1+z}}{{(1+z)}^2} \, \right]}^{D-3},
    \lb{gamidAD}
\ee
\be \gamid{G}{D}(z) = \left(\frac{D{H_0}^3}{32\pi c^3}\right)
    {\left(\frac{2cB}{H_0}\right)}^D \;
    {\left(\frac{1+z-\sqrt{1+z}}{1+z} \, \right)}^{D-3},
    \lb{gamidGD}
\ee
\be \gamid{L}{D}(z) = \left(\frac{D{H_0}^3}{32\pi c^3}\right)
    {\left(\frac{2cB}{H_0}\right)}^D \;
    {\left({1+z-\sqrt{1+z}} \, \right)}^{D-3},
    \lb{gamidLD}
\ee
\be \gamid{Z}{D}(z) = \left(\frac{D{H_0}^3}{32\pi c^3}\right)
    {\left(\frac{2cB}{H_0}\right)}^D \;
    {\left(\frac{z}{2}\right)}^{D-3}.
    \lb{gamidZD}
\ee
For the integral differential density given by Eq.\
(\ref{gestPW}), the expression in Eq.\ (\ref{doisgs}) allows
us to write the results as below,
\be \gamsid{A}{D}(z) = \left(\frac{3{H_0}^3}{32\pi c^3}\right)
    {\left(\frac{2cB}{H_0}\right)}^D \;
    {\left[\frac{1+z-\sqrt{1+z}}{{(1+z)}^2} \, \right]}^{D-3},
    \lb{gamsidAD}
\ee
\be \gamsid{G}{D}(z) = \left(\frac{3{H_0}^3}{32\pi c^3}\right)
    {\left(\frac{2cB}{H_0}\right)}^D \;
    {\left(\frac{1+z-\sqrt{1+z}}{1+z} \, \right)}^{D-3},
    \lb{gamsidGD}
\ee
\be \gamsid{L}{D}(z) = \left(\frac{3{H_0}^3}{32\pi c^3}\right)
    {\left(\frac{2cB}{H_0}\right)}^D \;
    {\left({1+z-\sqrt{1+z}} \, \right)}^{D-3},
    \lb{gamsidLD}
\ee
\be \gamsid{Z}{D}(z) = \left(\frac{3{H_0}^3}{32\pi c^3}\right)
    {\left(\frac{2cB}{H_0}\right)}^D \;
    {\left(\frac{z}{2}\right)}^{D-3}.
    \lb{gamsidZD}
\ee

When $D=3$, the definition (\ref{oh}) is fulfilled in view of
Eqs.\ (\ref{diffPW}) and (\ref{gestPW}) above and, therefore,
this choice of dimension clearly corresponds to OH. From this
follows an interesting result. We saw in Sect.\ \ref{eds} that
densities constructed using the galaxy area distance $\dg$ in an
EdS cosmology produces a model having both SH and OH. This
property allows us to find the constant $B$. For $D=3$, we can
equate Eq.\ (\ref{gamasg}) to Eq.\ (\ref{gamsidGD}) and,
considering Eq.\ (\ref{mu0}), we obtain the following result,
\be B={\left(\frac{{H_0}^2}{2M_gG}\right)}^{1/3}.
    \lb{B}
\ee
This is valid as long as the geometrical part of the model is
given by EdS spacetime. Thus, each adopted metric corresponds to 
a different value of the constant $B$, whose dimension is
${[\,\mbox{length unit}\,]}^{-1}$.

\subsection{Case of $D=3$}\lb{d3}

As seen above, this is the condition for the existence of OH and
holds for \textit{all} distance definitions (see Eq.\ \ref{gestPW}).
This case reduces Eq.\ (\ref{doisgs}) to the following
simple expression,
\be \gamsid{\sty i}{3} = \gamid{\sty i}{3} = 
    \frac{3 {H_0}^2}{8\pi M_g G}.
    \lb{special}
\ee
Expressions for the number counts $\NID{A}{3}(z)$,
$\NID{G}{3}(z)$, $\NID{L}{3}(z)$, and $\NID{Z}{3}(z)$ are
respectively obtained from Eqs.\ (\ref{Nad}),
(\ref{Ngd}), (\ref{Nld}), and (\ref{Nzd}). Recalling Eq.\
(\ref{neds}), we verify that $\NID{G}{3}(z)=N^{\mathrm{\ssty
EdS}}(z)$. These functions are plotted in Figs.\ \ref{fig2}
and \ref{fig34}, where the caption of the former Figure discusses
that only $\NID{G}{3}$ shows both SH and OH, although all
number-counting functions are observationally-homogeneous by
construction. 
%\newpage
\begin{figure}
\centering
\vspace{-5.cm}
\hspace{-2.cm} \includegraphics[width=12.5cm]{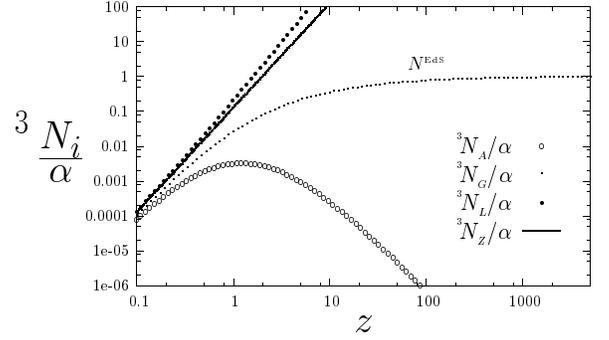}
%\vspace{0.6cm}
%\input{f3.tex}
\vspace{-8.5cm}
\caption{Number counts for the case $D=3$ in Eqs.\ (\ref{Nad}),
        (\ref{Ngd}), (\ref{Nld}), and (\ref{Nzd}). By definition these
        counts are \textit{all} observationally-homogeneous in 
        \textit{all} values of $z$, as can be seen from Eq.\
        (\ref{gestPW}). The four expressions start similarly, that
        is, close to the spatially-homogeneous case $\NID{G}{3}=
        N^\mathrm{\ssty EdS}$, but as $z$ increases deviations start
        to occur. At $z=0.5$ these deviations are significant.
        The only expression that exhibits both OH and SH is the one
        obtained from the galaxy area distance $\dg$. Since $\NID{G}
        {3}(z)$ and $\NID{L}{3}(z)$ have higher counts than
        $N^\mathrm{\ssty EdS}$ they should be spatially
        \textit{in}homogeneous. The same is also true for $\NID{A}{3}
        (z)$, although in this case the counts are less than the EdS
        one with SH. One can also easily verify that in the asymptotic
        limit of the big bang singularity hypersurface the following
        results hold,
        $\lim\limits_{z \rightarrow \infty} \NID{A}{3}=0$,
        $\lim\limits_{z \rightarrow \infty} \NID{G}{3}=\alpha$,
        $\lim\limits_{z \rightarrow \infty} \NID{L}{3}=\infty$,
        $\lim\limits_{z \rightarrow \infty} \NID{Z}{3}=\infty$.
        }
    \lb{fig2}
\end{figure}
\begin{figure}
\centering
\vspace{-5.cm}
\hspace{-2.cm} \includegraphics[width=12.5cm]{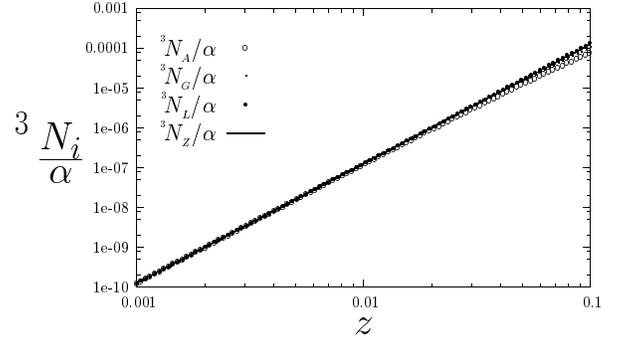}
%\vspace{0.9cm}
%\input{f4.tex}
\vspace{-8.5cm}
\caption{Graph also showing the number counts for the case $D=3$ in
         Eqs.\ (\ref{Nad}), (\ref{Ngd}), (\ref{Nld}), and (\ref{Nzd}),
	 but at small redshifts compared to the previous figure.
	 Clearly only at $z \approx 0.1$, the differences between the
         number counts constructed with the various distance definitions
	 can be seen.}
        \lb{fig34}
\end{figure}

\subsection{Case of $D=2$}\lb{d2}

We have observational \textit{in}homogeneity by construction for
any $D<3$ since Sylos Labini et al.\ (1998) reported a value
close to $D=2$ for the fractal dimension of the distribution of
galaxies, this however provides a suitable choice for our toy
model. {By definition, this choice implies no SH and, therefore,
it might be argued that the Friedmann models, as well as their
distance definitions, are not valid for any $D<3$. We emphasize
that our aim is not to validate completely the standard cosmology,
but to determine an unambiguous answer to the following question.
\textit{If} the number counting produced by assuming $D=2$ in
Eq.\ (\ref{nd}) \textit{corresponded to true observations}, can
we conclude, by applying a FLRW framework, that the galaxy distribution
does not follow the standard cosmology? In other words, using
galaxy counts simulated by Eq.\ (\ref{nd}) with $D=2$, is it
possible by employing procedures based on the standard
cosmological model to conclude with certainty that this number count
could not possibly be an observationally-homogeneous galaxy
distribution? This is a very important question as this approach was
taken by many studies found in the literature, that is, the framework
given by the standard cosmology was used to determine if the observed
data was consistent with this model. Although most who carry out these
studies implicitly assume that this is possible, as we show below, our
results indicate that ambiguities in interpreting observations within
the standard model framework still remain. As noted by Joyce et al.\
(2000), \textit{straightforward interpretations of FLRW standard cosmologies
are problematic; they include the interpretation that the isotropy of
microwave background radiation implies that observationally-inhomogeneous
matter distributions are impossible.}

We proceed and assume that $D=2$ in Eqs.\ (\ref{Nad}) to
(\ref{Nzd}). Considering the definitions provided in Eqs.\ (\ref{B})
and (\ref{alpha}), we obtain the following results,
\be {\; \NID{A}{2}}(z)={\alpha}^{2/3} \;
    {\left[\frac{1+z-\sqrt{1+z}}{{(1+z)}^2} \, \right]}^2,
    \lb{Nad2}
\ee
\be {\; \NID{G}{2}}(z)={\alpha}^{2/3} \;
    {\left(\frac{1+z-\sqrt{1+z}}{1+z} \, \right)}^2,
    \lb{Ngd2}
\ee
\be {\; \NID{L}{2}}(z)={\alpha}^{2/3} \;
    {\left({1+z-\sqrt{1+z}} \, \right)}^2,
    \lb{Nld2}
\ee
\be {\; \NID{Z}{2}}(z)={\alpha}^{2/3} \;
    {\left(\frac{z}{2}\right)}^2.
    \lb{Nzd2}
\ee
As can be seen in Fig.\ \ref{fig3}, the curves are similar
to those in Fig.\ \ref{fig2}, apart from the scale. Although
$\NID{G}{3}(z) \not = \NID{G}{2}(z)$, both functions converge to
a constant value in their asymptotic limits. As we show below,
this result has an interesting consequence.
\begin{figure}
\centering
\vspace{-5.cm}
\hspace{-2.cm} \includegraphics[width=12.5cm]{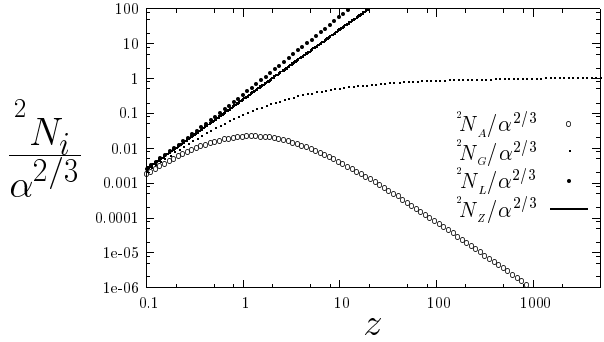}
\vspace{-8.5cm}
\caption{Number counts for the case of $D=2$. The functions show a
        behavior similar to those shown in figure \ref{fig2}. We note
        that $\NID{G}{2}$ tends to a constant value at its asymptotic
        limit. Indeed, the limits of the four functions are as follows,
        $\lim\limits_{z \rightarrow \infty} \NID{A}{2}=0$,
        $\lim\limits_{z \rightarrow \infty} \NID{G}{2}=\alpha^{2/3}$,
        $\lim\limits_{z \rightarrow \infty} \NID{L}{2}=\infty$, and
        $\lim\limits_{z \rightarrow \infty} \NID{Z}{2}=\infty$.
        }
    \lb{fig3}
\end{figure}
\begin{figure}
\centering
\vspace{-5.cm}
\hspace{-2.cm} \includegraphics[width=12.5cm]{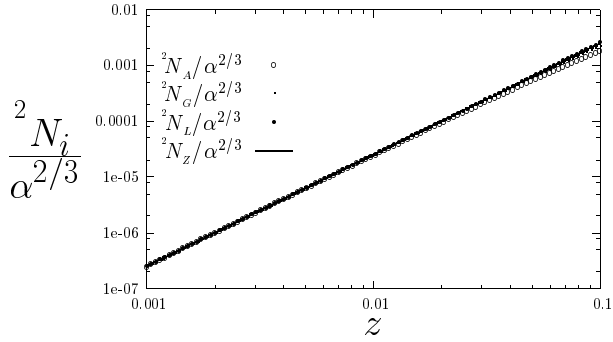}
\vspace{-8.5cm}
\caption{Number counts for the case where $D=2$ at very small
         redshifts. In a similar way to the graph shown in Fig.\
	 \ref{fig34}, only at $z \approx 0.1$ the various
	 distance measures begin to affect the number
	 counts. This range is, nevertheless, dependent on the
	 cosmological model, which means that if another cosmology
	 is adopted, deviations caused by the use of different
	 distance measures could possibly occur at redshifts
	 smaller than $z=0.1$.}

  \lb{fig56}
\end{figure}

The integral differential densities are calculated using Eqs.\
(\ref{gamsidAD}) to (\ref{B}), to be, 
\be \gamsid{A}{2}(z) = \mu_1 \;
    {\left[\frac{1+z-\sqrt{1+z}}{{(1+z)}^2} \, \right]}^{-1},
    \lb{gamsidAD2}
\ee
\be \gamsid{G}{2}(z) = \mu_1 \;
    {\left(\frac{1+z-\sqrt{1+z}}{1+z} \, \right)}^{-1},
    \lb{gamsidGD2}
\ee
\be \gamsid{L}{2}(z) = \mu_1 \;
    {\left({1+z-\sqrt{1+z}} \, \right)}^{-1},
    \lb{gamsidLD2}
\ee
\be \gamsid{Z}{2}(z) = \mu_1 \;
    {\left(\frac{z}{2}\right)}^{-1}.
    \lb{gamsidZD2}
\ee
where, 
\be \mu_1=\frac{3{H_0}^{7/3}}{8\pi c {(2M_g G)}^{2/3}}.
    \lb{mu1}
\ee
{We obtain the following power-series expansions of these
expressions, 
\be \gamsid{A}{2}(z) = \mu_1 \left(\frac{2}{z}+\frac{7}{2}+
    \frac{11}{8}z-\frac{1}{16}z^2 +  \ldots \right),
    \lb{gamsidAD2s}
\ee
\be \gamsid{G}{2}(z) = \mu_1 \left( \frac{2}{z}+\frac{3}{2}-
    \frac{1}{8}z+\frac{1}{16}z^2-  \ldots \right),
    \lb{gamsidGD2s}
\ee
\be \gamsid{L}{2}(z) = \mu_1 \left(\frac{2}{z}-\frac{1}{2}+
    \frac{3}{8}z-\frac{5}{16}z^2+  \ldots \right).
    \lb{gamsidLD2s}
\ee
We note that for small redshifts the first terms of the 
series above dominate. Considering the approximation provided
by Eq.\ (\ref{gamsidZD2}), we therefore have that
$\gamsid{A}{2}$, $\gamsid{G}{2}$ and $\gamsid{L}{2}$
become equal to $\gamsid{Z}{2}$.}

The functions above are plotted versus the redshift in Figs.\
\ref{fig4} and \ref{fig78}. We can see clearly in Fig.\ \ref{fig4}
that, although we started with an observationally-\textit{in}homogeneous
model, if we use the galaxy area distance $\dg$, the density, as a
function of redshift, is constant for $z>10$. In other words,
\textit{even a model that is observationally-inhomogeneous by
construction appears to become observationally-homogeneous at higher
$z$, if the density is built using the appropriate distance measure,
in this case $\dg$}. Both $\dl$ and $\dz$ produce observational
inhomogeneity for any $z$, which is reproduced as a power law decay.
The density constructed with the area distance $\da$ has an odd
behavior while starting to decay as a power law at small redshifts,
but at $z \approx 1$ this decay turns into an increase.

It is clear from these results that developing an
observationally-inhomogeneous density that decreases as $z$ increases
is no guarantee that it will remain so for all $z$. Clearly the use of
the various distance measures creates too many ambiguities, which prevent
definitive conclusions being made about the behavior of the large-scale
galaxy distribution in the Universe.\footnote{We note that a further
source of ambiguity in defining OH is that here it is defined
in terms of a density that remains unchanged when the generic
\textit{distance} $d_i$ changes, whereas we might possibly conceive a
definition of OH in terms of densities that remain unchanged for
different values of the \textit{redshift}.}

\begin{figure}
\centering
\vspace{-5.cm}
\hspace{-2.cm} \includegraphics[width=12.5cm]{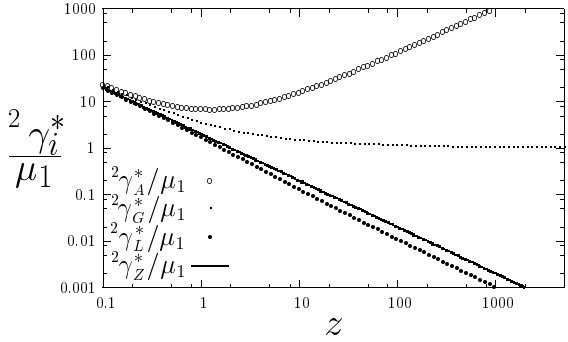}
\vspace{-8.5cm}
\caption{Graph of the integral differential densities versus the
        redshift for the case of $D=2$. It is easy to show that
	the following asymptotic limits hold,
        $\lim\limits_{z \rightarrow \infty} \gamsid{A}{2}=\infty$,
        $\lim\limits_{z \rightarrow \infty} \gamsid{G}{2}=\mu_1$,
        $\lim\limits_{z \rightarrow \infty} \gamsid{L}{2}=0$, and
        $\lim\limits_{z \rightarrow \infty} \gamsid{Z}{2}=0$.
        All densities have a power-law decay at small redshifts.
	However, at $z \approx 1$ the densities constructed with the
	redshift and luminosity distances continue to decay, whereas
	$\gamsid{A}{2}$ begins to change from a decay to an increasing
	behavior. More interestingly, the density constructed with the
	galaxy area distance $\gamsid{G}{2}$ begins to change from a
	power-law decay to a constant value at $z \approx 2.5$. This
	is a consequence of the fact that the number counting
	constructed with $\dg$ becomes constant at higher values of
	$z$ (see figure \ref{fig3}). Since $\NID{G}{3}$ tends to a
	constant value as the redshift increases, and since
	$\NID{G}{3}=N^\mathrm{\ssty EdS}$ (see figure \ref{fig2}),
	this suggests that $\gamsid{G}{2} \;$ becomes spatially
	\textit{homogeneous} for $z > 10$. This occurs despite the
	fact that this density is observationally \textit{in}homogeneous
	by construction. This result therefore suggests that for
	$D=2$ the integral differential density constructed with
	the galaxy area distance $\dg$ is \textit{not} observationally
	and spatially homogeneous for $z<10$, but it seems to turn
	into both for $z>10$. This is a simple and clear example
	showing that the use of different distance measures in the
	characterization of cosmological densities may lead to
	significant ambiguities in reaching conclusions about the
	behavior of the large-scale galaxy distribution.
        }
    \lb{fig4}
\end{figure}
\begin{figure}
\centering
\vspace{-5.cm}
\hspace{-2.cm} \includegraphics[width=12.5cm]{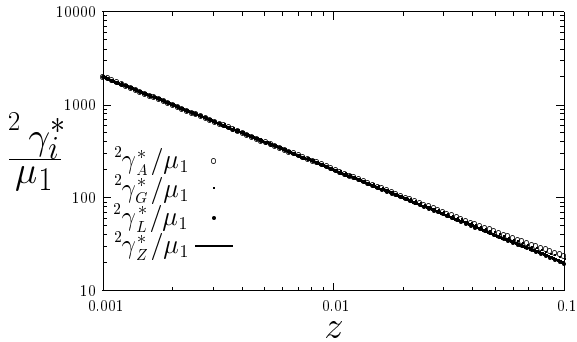}
\vspace{-8.5cm}
\caption{Graph of the integral differential densities versus the
        redshift for the case of $D=2$ at very small redshifts.
	Only when $z \approx 0.1$, this density starts to be affected
	by the different cosmological distance definitions.}
	\lb{fig78}
\end{figure}

\section{Behavior of the dimension D in a spatially
            homogeneous case}\lb{DSH}

We have so far assumed constant values for the dimension $D$. However,
it is interesting to study the behavior of this dimension as one
approaches the Big Bang singularity hypersurface. Inasmuch as it is
well known that at those very early times curvature effects become
negligible, studying the behavior of $D$ as $z \rightarrow \infty$ in
the EdS model should shed some light on its general behavior. To do
so, we proceed as follows.

In the EdS model, the generalized number-distance relation given by Eq.\
(\ref{nd}) may be written as below if we consider the number counting
given by Eq.\ (\ref{neds}),
\be { \left( B \, d_i \right) }^D = \alpha \; 
    { \left( \frac{1+z-\sqrt{1+z}}{1+z} \right) }^3.
    \lb{dsh}
\ee
These are in fact four equations, one for each distance definition
indicated by the index $i={\sty A}, {\sty G}, {\sty L}, {\sty Z}$.
We define the EdS distance measures as follows,
\be d_i(z)=\frac{2c}{H_0}f_i(z),
    \lb{fi}
\ee
where $f_i(z)$ is a function given by each distance definition provided
in Eqs.\ (\ref{da}), (\ref{dg}), (\ref{dl}), and (\ref{dz1}).
Considering the definitions provided by Eqs.\ (\ref{alpha}) and (\ref{B}),
the four equations given by the expression (\ref{dsh}) may be rewritten
in terms of the dimension $D$, as can be seen below,
\be D_i=3\left( \frac{\ln \alpha^{1/3} + \ln f_{\ssty G}}{\ln
        \alpha^{1/3} + \ln f_i} \right).
    \lb{Di}
\ee
The following result comes directly from this Eq.\ (\ref{Di}),
\be D_{\ssty G}=3.
    \lb{Dg}
\ee
This value for the fractal dimension should not come as a surprise
since the density defined with the galaxy area distance $\dg$
remains constant, that is, observationally homogeneous for all $z$
in the EdS cosmology (see figure \ref{fig1}). The other three
expressions for the dimension $D$ in each distance measure yield,
\be D_{\ssty A}=3\left\{ \frac{\ln \alpha^{1/3} +
       \ln \left( \displaystyle \frac{1+z-\sqrt{1+z}}{1+z} \right)}{\ln
        \alpha^{1/3} + \ln \left[ \displaystyle
        \frac{1+z-\sqrt{1+z}}{(1+z\,)^{\,2}} \right]} \right\},
    \lb{DA}
\ee
\be D_{\ssty L}=3\left[ \frac{\ln \alpha^{1/3} +
       \ln \left( \displaystyle \frac{1+z-\sqrt{1+z}}{1+z} \right)}{\ln
        \alpha^{1/3} + \ln \left( 1+z-\sqrt{1+z} \right) } \right],
    \lb{DL}
\ee
\be D_{\ssty Z}=3\left[ \frac{\ln \alpha^{1/3} +
       \ln \left( \displaystyle \frac{1+z-\sqrt{1+z}}{1+z} \right)}{\ln
        \alpha^{1/3} + \ln \displaystyle \left( \frac{z}{2} \right)} \right].
    \lb{DZ}
\ee
{Series expansions for the expressions above may be written
as follows,
\be D_{\ssty A}=\left[ 3+\left(\frac{3z}{\ln \alpha^{1/3} - \ln 2 +
   \ln z}\right)+ \ldots \right],
    \lb{DAs}
\ee
\be D_{\ssty L}=\left[ 3-\left( \frac{3z}{\ln \alpha^{1/3} - \ln 2 +
    \ln z} \right) + \ldots \right],
    \lb{DLs}
\ee
\be D_{\ssty Z}=\left[ 3- \left(\frac{3}{4}\right) \left( \frac{3z}
    { \ln \alpha^{1/3} - \ln 2 + \ln z  }  \right) + \ldots \right].
    \lb{DZs}
\ee
Clearly these results are valid for small, but nonzero, values of the
redshift. However,} we show that functions given by the Eqs.\ (\ref{DA}),
(\ref{DL}), and (\ref{DZ}) converge as the redshift vanishes. In fact,
we have the following results,
\be \lim\limits_{z \rightarrow 0} D_{\ssty A}=3,
    \lb{limDa}
\ee
\be \lim\limits_{z \rightarrow 0} D_{\ssty L}=3,
    \lb{limDl}
\ee
\be \lim\limits_{z \rightarrow 0} D_{\ssty Z}=3.
    \lb{limDz}
\ee
Again, these results are not surprising since, according to
Fig.\ \ref{fig1}, all densities tend to OH at very small
redshifts. The asymptotic limits of these functions at the
Big Bang are also found easily, yielding, 
\be \lim\limits_{z \rightarrow \infty} D_{\ssty A}=0,
    \lb{limDa1}
\ee
\be \lim\limits_{z \rightarrow \infty} D_{\ssty L}=0,
    \lb{limDl1}
\ee
\be \lim\limits_{z \rightarrow \infty} D_{\ssty Z}=0.
    \lb{limDz1}
\ee
We compare these results with Eq.\ (\ref{Dg}) and conclude
that the dimension $D$ constructed with the area distance $\da$,
luminosity distance $\dl$, and redshift distance $\dz$
corresponds to a vanishing fractal dimension at the Big Bang,
whereas that constructed with the galaxy area distance $\dg$
(or comoving distance) produce a finite nonzero dimension at
the Big Bang singularity hypersurface. That indicates that the
ambiguities arise when one considers all distance measures in
cosmology, in view of the fact that the value of the fractal
dimension at the Big Bang depends on the chosen cosmological
distance definition. These results are graphically shown in
Figs.\ \ref{fig5} and \ref{fig910}.
\begin{figure}
\centering
\vspace{-5.cm}
\hspace{-2.cm} \includegraphics[width=12.5cm]{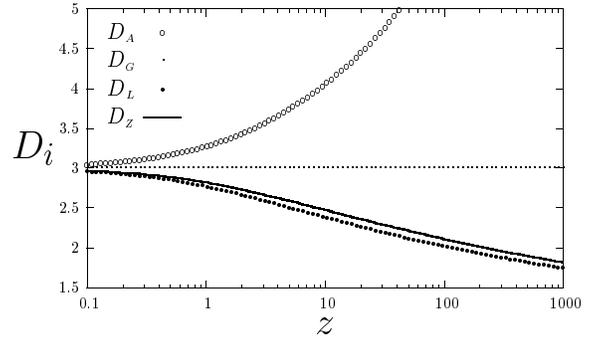}
\vspace{-8.5cm}
\caption{Graph of the dimension $D$ defined in terms of the four
        distance measures $\da$, $\dg$, $\dl$, and $\dz$ and plotted
        against the redshift in the spatially-homogeneous EdS
        cosmological model. Assuming $H_0=70$ km s$^{-1}$ Mpc$^{-1}$
        and $M_g=10^{11} M_\odot$, we have $\ln\alpha^{1/3}=9.6353$.
        For $z \ll 1$, all dimensions are equal to 3 (see Fig.\
	\ref{fig910} and Eqs.\ [\ref{Dg}], [\ref{limDa}],
	[\ref{limDl}], [\ref{limDz}]). For higher values of $z$,
	both $D_{\ssty L}$ and $D_{\ssty Z}$ decrease steadily and
	vanish as $z \rightarrow \infty$. The dimension $D_{\ssty
	A}$ constructed with the area distance $\da$ shows an odd
	behavior, initially increasing very rapidly well above 3
	for $z>0.1$. However, according to Eq.\ (\ref{limDa1}),
	it eventually vanishes at the Big Bang singularity
	hypersurface which implies that this function must experience
        dramatic changes. Indeed, it is discontinuous at
        $z \approx 15170$, changing to negative values that increase
        towards zero. $D_{\ssty G}$ remains constant for all
        redshifts. This plot is a different way of presenting the
        results in Fig.\ \ref{fig1}. It is clear from this graph that
        although the EdS cosmology is spatially homogeneous, it may or
        may not be observationally homogeneous depending on the distance
	measure adopted for analyzing the behavior of the density
	in this model.
        }
    \lb{fig5}
\end{figure}
\begin{figure}
\centering
\vspace{-5.cm}
\hspace{-2.cm} \includegraphics[width=12.5cm]{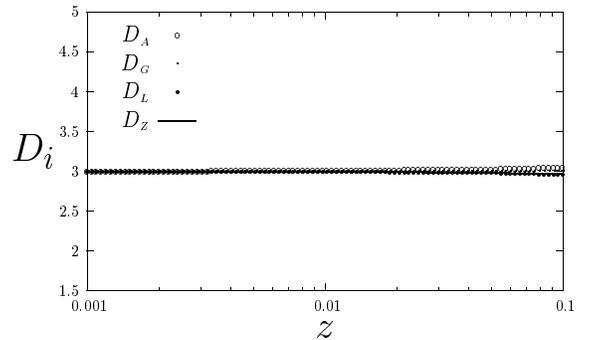}
\vspace{-8.5cm}
\caption{Graph of the dimension $D$ versus the redshift at very small
         values of $z$. Note that only when $z \approx 0.1$ that
	 deviations from the constant values $D=3$ start to appear
	 due to the definition of $D(z)$ in terms of the four
         distance measures.}
    \lb{fig910}
\end{figure}

\section{Number counts and magnitudes}\lb{mags}

{The results discussed in the previous sections show that
if we use the framework of the EdS cosmology, only when
we have $z>0.1$ it does become theoretically possible to detect
the distinction between SH and OH. If we depart from EdS
cosmology and use, for instance, an open FLRW model the
redshift ranges where this distinction becomes detectable
could also change. Indeed, by means of differential
densities calculated from the CNOC2 galaxy redshift survey
data in the range $0.05 \le z \le 1$ in a FLRW model with
$\Omega_{m_0}=0.3$ and $\Omega_{\Lambda_0}=0.7$, we are able
to detect such a distinction for $z \approx 0.1$ (see A07)
but, to do so we need to obtain indirectly the
number counts by employing a method capable of calculating
them from the galaxy luminosity function (LF). Presently
the LF is evaluated from detailed observations of the
apparent magnitude of galaxies and such magnitude
measurements can go to redshifts much higher than the
unity. To extract number counts from the LF, we
require a relativistic theory connecting the theoretical
aspects of a cosmological model with the usual procedures
carried out by astronomers when quantifying galaxy
catalogues. Ribeiro \& Stoeger (2003) developed such a
method, A07 applied to the CNOC2 galaxy survey, and
Iribarrem, Ribeiro \& Stoeger (2008, in preparation) applied
to the FORS Deep Field redshift survey. These articles dealt
with issues such as source evolution and K-correction (see
also Ribeiro 2002) and the reader interested in those topics
is referred to these articles since it is beyond the scope of
this work to present a detailed discussion of these issues.

Despite this, a simpler discussion about the relationship
between number counts and magnitudes can be presented in the
context of this paper. We define the bolometric \textit{apparent
magnitude} $m$ to be given by the following Eq.,
\be m=-2.5 \log \left( \frac{L}{4 \pi {\dl}^2} \right)
      + \mathrm{const.},
    \lb{m}
\ee
where $L$ is the intrinsic bolometric luminosity of a
cosmological source, assumed point like (Ribeiro 2002),
and the constant is due to the calibration of the magnitude
system. Since by definition the bolometric \textit{absolute
magnitude} $M$ is defined to be the apparent magnitude of a
source located at a distance of 10 pc, the \textit{distance
modulus} is defined as follows,
\be \mu \equiv m-M = 5 \log \dl + 25.
    \lb{dm}
\ee
In this equation, the luminosity distance is measured in Mpc.
Thus, if astronomical observations provide measurements of
the pair $(m;z)$ for cosmological sources, which is common for
sources of redshifts higher than unity, then, by assuming a
cosmological model, we have the function $\dl=\dl(z)$ from which
we can calculate $L$ using Eq.\ (\ref{m}) and then derive $M$ and
$\mu$ for these sources. Therefore, the distance modulus is
another way of representing observed magnitudes and
redshifts of cosmological sources.

Our aim is to relate the number counts given for each
distance adopted in this paper to the distance modulus. This
can be done if we use the reciprocity theorem provided by Eq.\
(\ref{eth}) to write $\nid{A}$, $\nid{G}$, and $\nid{L}$, as given
by Eq.\ (\ref{nd}), in terms of the luminosity distance
and then use Eq.\ (\ref{dm}) to write the final
expressions in terms of the distance modulus. The results may
be written as follows,
\be \nid{A} = { \left[ B \frac{10^{0.2(\mu-25)}}{{(1+z)}^2}
              \right] }^D,
    \lb{dnamu}
\ee
\be \nid{G} = { \left[ B \frac{10^{0.2(\mu-25)}}{(1+z)}
              \right] }^D,
    \lb{dngmu}
\ee
\be \nid{L} = { \left[ B \, 10^{0.2(\mu-25)} \right] }^D,
    \lb{dnlmu}
\ee
\be \nid{Z} = { \left[ \frac{cB}{H_0}z \right] }^D,
    \lb{dnzmu}
\ee
where the last Eq.\ is the result of simply substituting
the definition given by Eq.\ (\ref{dz}) for the redshift
distance, directly into Eq.\ (\ref{nd}) when taking $i=z$.

To express the number counts above only in terms
of the distance modulus, we require the function $z=z(\mu)$
and to derive it we need to adopt a cosmological model as we
did in the previous sections. In terms of observations, a
cosmological model is required from the beginning of this approach,
otherwise it would be impossible to find absolute magnitudes
and, therefore, the distance moduli. As in previous sections, we
adopt the EdS cosmology, which corresponds to using
the luminosity distance as given in Eq.\ (\ref{dl}).
Recalling that $\dl=0$ when $z=0$ Eq.\ (\ref{dl}) can
be inverted to produce $z=z(\dl)$ and, after considering
Eq.\ (\ref{dm}), we finally obtain the function $z(\mu)$
given by,
\be 1+z= \frac{1}{2}+ \frac{H_0}{2c}10^{0.2(\mu-25)}+ 
        \sqrt{ \frac{H_0}{2c}10^{0.2(\mu-25)}+ \frac{1}{4}}
        \; \; .
    \lb{zmu}
\ee
We note that Eqs.\ (\ref{dl}) and (\ref{dm}) imply that the
small redshifts interval $0.001 \le z \le 0.1$ corresponds 
to $4.3 \, \mathrm{Mpc} \le \dl \le 439 \, \mathrm{Mpc}$ and
$28.2 \le \mu \le 38.2$, if we assume that
$H_0=70$~km~s$^{-1}$~Mpc$^{-1}$.

For number counts to correspond to OH, we must choose $D=3$, which
implies that the EdS cosmology number counts distribution is
derived as in Sect.\ \ref{d3}, that is, is given by
$N^\mathrm{\ssty EdS}(\mu)= \NID{G}{3} (\mu)$. Thus, only
the expression $\NID{G}{3} (\mu)$ produces a number count
distribution that is both OH and SH whereas the remaining
functions $\NID{A}{3}(\mu)$, $\NID{L}{3}(\mu)$, and $\NID{Z}{3}
(\mu)$ are OH, but are not SH. Figures \ref{fig-11} and
\ref{fig-12} show graphs of these functions where it is
clear that they are similar to those plotted in Figs.\
\ref{fig2} and \ref{fig34}. Therefore, although these
four expressions are built with a geometrical part
consistent with EdS cosmology, this does not imply that
the counts will be SH and OH, showing again the ambiguous
nature of these expressions as far as ``homogeneity'' is
concerned. 
\begin{figure}
\centering
\vspace{-5.cm}
\hspace{-2.cm} \includegraphics[width=12.5cm]{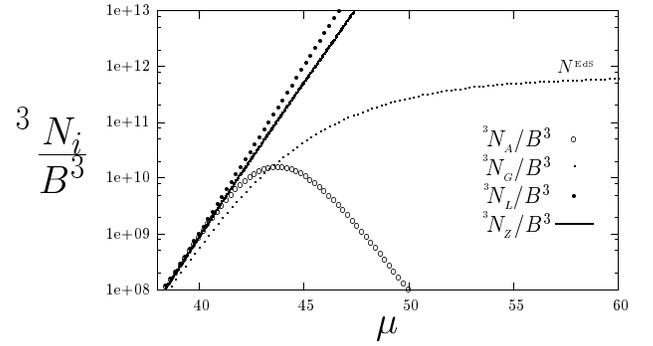}
\vspace{-8.5cm}
\caption{This graph shows number counts versus distance modulus
         for the case $D=3$. The constant $B$ is given by Eq.\
	 (\ref{nd}) which, for the EdS cosmology, turns out to be
	 equal to Eq.\ (\ref{B}). The results are very similar
	 to the ones shown in figure \ref{fig2}, meaning that the
	 same conclusions reached there apply here.}
    \lb{fig-11}
\end{figure}
\begin{figure}
\centering
\vspace{-5.cm}
\hspace{-2.cm} \includegraphics[width=12.5cm]{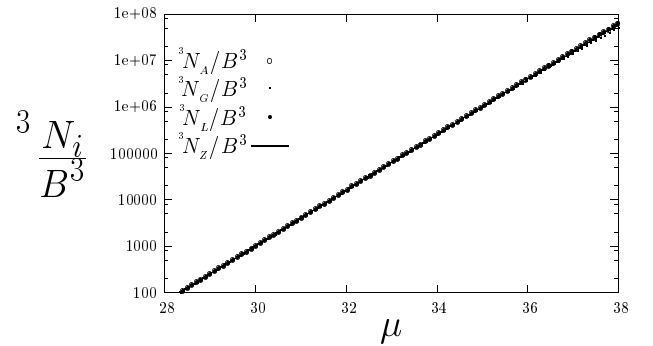}
\vspace{-8.5cm}
\caption{This is the same plot shown in figure \ref{fig-11}, but
         with a distance modulus range equivalent to small redshifts
	 interval.}
    \lb{fig-12}
\end{figure}

}

\section{The relativity of observational homogeneity}\lb{rel}

We have seen how the concept of homogeneity applied to cosmological
models is prone to ambiguities. Attempting to distinguish between
spatial and observational homogeneities is a means of diminishing the
ambiguities that, as seen above, have not been eliminated; this is
because using various distance measures to calculate cosmological
densities is still a source of ambiguity. At this point we recall 
some well known concepts that may help to clarify the physical
interpretation of the effects discussed above.

According to the reciprocity theorem given by Eq.\ (\ref{eth}),
all distance
definitions discussed above become equal at $z=0$. This means that
if a signal such as pulses emitted at unit time intervals are
emitted at the rest frame of the source and an observer measures
the rate of change of the same signal, these rates of change are,
by definition, the redshift $z$. In particular, the observed
frequencies $\nu$ of light or radio waves are related to $z$ as
$1+z={\nu_{_{\mbox \tiny \rm emi}}}/{\nu_{_{\mbox \tiny \rm obs}}}.$
We can think of this as a {\it time dilation} effect (Ellis 1971).
So, if a proper time interval $dt$ is observed to elapse between
particular signals, then
\begin{equation}
  \frac{dt_{\mbox \tiny \rm emitted}}{dt_{\mbox \tiny \rm observed}}
  = \frac{\nu_{\mbox \tiny \rm emitted}}{\nu_{\mbox \tiny \rm
  observed}}=1+z.
  \label{red}
  \end{equation}
This relationship is true regardless of the separation of emitter and
observer and implies that the difference between two distance
measures in cosmology can be thought to be, in effect, a result of
the time dilation between emitter and observer located in different
reference frames in relative motion with one another. So, we can
consider Eq.\ (\ref{eth}) to be produced by the relativity of
time intervals. Inasmuch as the observable distances discussed in
the previous sections are used to build observational densities,
then we conclude that the concept of OH as defined in Eq.\
(\ref{oh}) must also be relative. This means that similarly to the
concept of a cosmological distance, we can talk about the relativity
of OH and, therefore, we must abandon the notion of a ``true'', or
unique, homogeneity of the observable Universe. In a similar way to
the statement of McVittie (1974) about cosmological distances, we
conclude that measuring the possible homogeneity of the large-scale
structure in the Universe depends on circumstances, that is, on the
method of measurement.

The reasoning presented above can then help us to understand the
limitations of the generic concept of homogeneity widely used in
cosmology. It has its origins in the assumption of the maximal
spatial isotropy in the Riemannian spacetime manifold, which then
follows, as a mathematical result, that a perfect fluid cosmology
metric ends up with its fluid variables (density and pressure)
being time dependent only (Stephani et al.\ 2003, pp.\ 173, 210-212; 
Weinberg 1972, pp.\ 403, 412-415). This means that the local
density $\rho$ that appears in the right hand side of Einstein's
field equations becomes a function of the time coordinate only.
Hence, a spatially-isotropic spacetime is, by mathematical
requirement, spatially homogeneous as well. Due to this widely
known result, it is usual to refer to the standard FLRW family of
cosmological models as being characterized by isotropy and
homogeneity. The adjective ``spatial'' is often dropped from
appearing in front of the term homogeneity when the most basic
features of the standard cosmology are described (e.g., see
Peacock 1999, p.\ 65).

Such an economy of language could, perhaps, have been thought 
harmless, but as a side effect it has in practice created a
simplistic, but wrong, impression that all types of densities
that can be derived in these cosmologies must also be
homogeneous, that is, eventually become a constant value.
As discussed in the previous sections, in contrast to this
simplistic view it is possible to define densities in standard
cosmologies that have different types of physical
homogeneity.\footnote{Note that in this paper the term homogeneity
was used with a physical meaning related to the average density,
which can, in principle, be empirically determined, directly, or
indirectly, by means of astronomical observations. Therefore,
homogeneity has a wider meaning than the strict mathematical
sense of spacetimes admitting isometries due to groups of motions
(Stephani et al.\ 2003, pp.\ 157, 171).} The concept of OH is
therefore fundamentally different from the concept of SH and,
therefore, it is a misleading use of language to call the
standard FLRW family of cosmological models simply isotropic and
homogeneous. They are in fact isotropic and spatially homogeneous
and either can or cannot be observationally homogeneous as well.

%______________________________________________________________

\section{Conclusion}\lb{conclu}

In this paper we have presented an analysis of the physical
consequences of the distinction between the usual concept of
spatial homogeneity, as defined by the Cosmological Principle,
and the concept of observational homogeneity. This distinction
is based on calculating observational areas, volumes and
densities with four cosmological distance measures
$d_i \; (i={\sty A}, {\sty G}, {\sty L}, {\sty Z} \,)$,
namely the area distance $\da$, the galaxy area distance $\dg$,
the luminosity distance $\dl$, and the redshift distance $\dz$.
Our aim was to simulate number counts as if they were actual
observations. To do so in a general way, we have
adopted the generalized number-distance relation $\Nid =
{ \left( B \, d_i \right) }^D$ to obtain the
differential density $\gammaid$ and the integral differential
density $\gestid$, where the latter is the observed radial
number density $\anid$. In this way, these densities become a
function of the fractal dimension $D$. We then reviewed the
results of Ribeiro (2001b, 2005) and Albani et al.\ (2007),
where those equations were applied to the Einstein-de Sitter
cosmological model and the open FLRW cosmology with
$\Omega_{m_0} =0.3$, $\Omega_{\Lambda_0}=0.7$, concluding
that a spatially-homogeneous cosmology does not necessarily
possess observational homogeneity. These features are only
present if the galaxy area distance $\dg$, which in EdS
cosmology is equivalent to the comoving distance apart from a
constant, is used to calculate both differential densities. 

Models with and without observational homogeneity by construction
were studied by means of setting $D=3$ and $D=2$ respectively in
the generalized number-distance relation. It was found that
models with $D=3$ do not seem to remain spatially homogeneous as
well. The only exception appears to be when one adopts the galaxy
area distance $\dg$. Models with $D=2$ were developed to be
observationally inhomogeneous, although the integral differential
density was constructed with the galaxy area distance $\dg$, which
when plotted versus redshift shows a power-law decay for $z<2.5$
that, after a transition, turns into a constant value for $z>10$.
We have also studied the behavior of the dimension $D$ for
the spatially-homogeneous EdS cosmology, showing that it tends
to $D=3$ as $z \rightarrow 0$ for all distance measures, tends to
$D=0$ as $z \rightarrow \infty$ for $\da$, $\dg$, and $\dl$,
but remains $D=3$ for $\dg$ at the Big Bang singularity
hypersurface. {Finally, we have also studied functions of
number counts versus distance modulus with the various distance
definitions and reached conclusions similar to models with
$D=3$ and functions of number counts versus redshift.} The paper
finishes with a conceptual discussion arguing that due to the
relativity of time intervals for pulses emitted and observed at
different reference frames, and in view of the reciprocity
theorem linking various cosmological distances by means of
$(1+z)$ factors, we can conclude that the concept of observational
homogeneity should also be relative.

To end this paper, it is important to emphasize that the conceptual
distinction discussed above between different types of homogeneity
in the standard cosmological model is fundamental and has important
consequences for observational cosmology. In view of the fact that
such a distinction is not generally recognized in the literature of
observational cosmology, it is our opinion that it should be
considered by all those who empirically probe the possible
observational homogeneity of the large-scale distribution of galaxies
in the Universe.

%______________________________________________________________

\begin{acknowledgements}
  We thank M. Montuori for discussions that motivated this work.
  {We are also grateful to a referee for interesting remarks,
  which improved the paper.} One of us (LJRL) acknowledges the financial
  support from CAPES Foundation.
\end{acknowledgements}


\begin{thebibliography}{}

\bibitem[2001]{amr01} Abdalla, E., Mohayaee, R., \& Ribeiro, M.B.\ 2001,
        Fractals 9, 451, arXiv:astro-ph/9910003v4
\bibitem[2007]{airs07} Albani, V.V.L., Iribarrem, A.S., Ribeiro, M.B.,
        \& Stoeger, W.R. 2007, Astrophys.\ J.\ 657, 760,
	arXiv:astro-ph/0611032v1 \textbf{(A07)}
\bibitem[1970]{dV70} de Vaucouleurs, G.\ 1970, Science 167, 1203
\bibitem[1971]{e71} Ellis, G.F.R.\ 1971, General Relativity and
        Cosmology, Proc.\ Int.\ School of Phys.\ ``Enrico Fermi'',
	ed.\ R.K.\ Sachs (Academic Press, New York), 104
\bibitem[2007]{e07} {Ellis, G.F.R.\ 2007, Gen.\ Rel.\ Grav.\
        39, 1047}
\bibitem[1933]{eth33} Etherington, I.M.H.\ 1933, Phil.\ Mag.\ 15, 761;
        {reprinted in Gen.\ Rel.\ Grav.\ 39 (2007) 1055-1067}
\bibitem[2005]{gsjp05} Gabrielli, A., Sylos Labini, F., Joyce, M., \&
        Pietronero, L.\ 2005, Statistical Physics for Cosmic Structures
        (Springer, Berlin)
\bibitem[2005]{hebbbgs05}Hogg, D.W., Eisenstein, D.J., Blanton, M.R.,
        Bahcall, N.A., Brinkmann, J., Gunn, J.E., \& Schneider, D.P.
        2005, Astrophys.\ J.\ 624, 54, arXiv:astro-ph/0411197v1
\bibitem[2005]{jmst05} Jones, B.J.T., Mart\'{\i}nez, V.J., Saar, E.,
        \& Trimble, V.\ 2005, Rev.\ Mod.\ Phys.\ 76, 1211,
	arXiv:astro-ph/0406086v1
\bibitem[1999]{jmsp99} Joyce, M., Montuori, M., Sylos Labini, F., \&
        Pietronero, L.\ 1999, A\&A 344, 387, arXiv:astro-ph/9805126v3
\bibitem[2000]{jamps00} {Joyce, M., Anderson, P.\ W., Montuori, M.,
        Pietronero, L., \& Sylos Labini, F.\ 2000, Europhys.\ Lett.\
	50, 416, arXiv:astro-ph/0002504v1}
\bibitem[2005]{jsgmp05} Joyce, M., Sylos Labini, F., Gabrielli, A.,
        Montuori, M., \& Pietronero, L.\ 2005, A\&A 443, 11,
	arXiv:astro-ph/0501583v2
\bibitem[1974]{m74} McVittie, G.C.\ 1974, Quart.\ J.\ Royal Astr.\
        Soc.\ 15, 246
\bibitem[1999]{p99} Peacock, J.A.\ 1999, Cosmological Physics (Cambridge
        University Press)
\bibitem[1987]{p87} Pietronero, L.\ 1987, Physica A 144, 257
\bibitem[2006]{pk06} Pleba\'{n}ski, J., \& Krasi\'{n}ski, A.\ 2006, An
        Introduction to General Relativity and Cosmology (Cambridge
        University Press)
\bibitem[1992]{rib92} Ribeiro, M.B.\ 1992, On Modelling a Relativistic
        Hierarchical (Fractal) Cosmology by Tolman's Spacetime, Ph.D.\
	Thesis, Queen Mary and Westfield College, University of London
\bibitem[1992a]{rib92a} Ribeiro, M.B.\ 1992a, Astrophys.\ J.\ 388, 1 
\bibitem[1992b]{rib92b} Ribeiro, M.B.\ 1992b, Astrophys.\ J.\ 395, 29 
\bibitem[1993]{rib93} Ribeiro, M.B.\ 1993, Astrophys.\ J.\ 415, 469 
\bibitem[1994]{rib94} Ribeiro, M.B.\ 1994, in Deterministic Chaos in
        General Relativity, eds.\ D.W.\ Hobill, A.\ Burd, \& A.\
        Coley (Plenum, New York) 269
\bibitem[1995]{rib95} Ribeiro, M.B.\ 1995, Astrophys.\ J.\ 441, 477,
        arXiv:astro-ph/9910145v1
\bibitem[2001a]{rib01a} Ribeiro, M.B.\ 2001a, Fractals 9, 237,
        arXiv:gr-qc/9909093v2
\bibitem[2001b]{rib01b} Ribeiro, M.B.\ 2001b, Gen.\ Rel.\ Grav.\ 33,
        1699, arXiv:astro-ph/0104181v1 \textbf{(R01b)} 
\bibitem[2002]{rib02} {Ribeiro, M.B.\ 2002, Observatory 122,
        201, arXiv:gr-qc/9910014v2}
\bibitem[2005]{rib05} Ribeiro, M.B.\ 2005, A\&A 429, 65,
        arXiv:astro-ph/0408316v2 \textbf{(R05)}
\bibitem[1998]{rm98} Ribeiro, M.B., \& Miguelote, A.Y.\ 1998, Braz.\
        J.\ Phys.\ 28, 132, arXiv:astro-ph/9803218v1
\bibitem[2003]{rs03} {Ribeiro, M.B., \& Stoeger, W.R.\ 2003,
        Astrophys.\ J.\ 592, 1, arXiv:astro-ph/0304094v1}
\bibitem[1988]{s88} Sandage, A.\ 1988, Ann.\ Rev.\ Astron.\
        Astrophys.\ 26, 561
\bibitem[2003]{skmhh03} Stephani, H., Kramer, D., MacCallum, M.A.H.,
        Hoenselaers, C, \& Herlt, E.\ 2003, Exact Solutions of Einstein's
        Field Equations, 2nd ed. (Cambridge University Press)
\bibitem[1998]{smp98} Sylos Labini, F., Montuori, M., \& Pietronero,
        L.\ 1998 Phys.\ Rep.\ 293, 61, arXiv:astro-ph/9711073v1
\bibitem[1972]{w72} Weinberg, S.\ 1972, Gravitation and Cosmology
        (Wiley, New York)
\bibitem[1970]{w70} Wertz, J.R.\ 1970, Newtonian Hierarchical Cosmology,
        Ph.D.\ Thesis, University of Texas at Austin 
\bibitem[1971]{w71} Wertz, J.R.\ 1971, Astrophys.\ J.\ 164, 227
\bibitem[2005]{ybps05} Yadav, J., Bharadwaj, S., Pandey, B., \&
        Seshadri, T.R.\ 2005, Mon.\ Not.\ Roy.\ Astron.\ Soc.\ 364, 601,
        {arXiv:astro-ph/0504315v1}
\end{thebibliography}
\end{document}